\documentclass[runningheads]{llncs}
\usepackage[T1]{fontenc}
\usepackage{graphicx}
\begin{document}
\title{Estimating Population Burden of Stroke with an Agent-Based Model}
%
%
\author{Elizabeth Hunter \inst{1}\orcidID{0000-0002-1767-4744} \and
John D. Kelleher\inst{2,3}\orcidID{0000-0001-6462-3248} }
\authorrunning{E. Hunter and J.D. Kelleher}
%
\institute{Technological University Dublin, Dublin, Ireland \and
Trinity College Dublin, Dublin, Ireland
\email{elizabeth.hunter@tudublin.ie}\\
}
\maketitle              
\begin{abstract}
 Stroke is one of the leading causes of death and disability worldwide but it is believed to be highly preventable.  The majority of stroke prevention focuses on targeting high-risk individuals but its is important to understand how the targeting of high-risk individuals might impact the overall societal burden of stroke.  We propose using an agent-based model that follows agents through their pre-stroke and stroke journey to assess the impacts of different interventions at the population level.  We present a case study looking at the impacts of agents being informed of  their stroke risk at certain ages and those agents taking measure to reduce their risk.  The results of our study show that if agents are aware of their risk and act accordingly we see a significant reduction in strokes and population DALYs. The case study highlights the importance of individuals understanding their own stroke risk for stroke prevention and the usefulness of agent-based models in assessing the impact of stroke interventions.  

\end{abstract}
\section{Introduction}

In 2019, stroke was the second leading cause of death worldwide and the third cause of death and disability combined \cite{feigin2022world}.  There are two main prevention methods for addressing stroke population wide strategies and high-risk strategies. Population wide strategies focus on reducing the prevalence of risk associated behaviours at a population level for example reducing the rates smoking or alcohol intake across populations, whereas high-risk strategies target interventions at those individuals who have been deemed to have high individual risk. Although it is believed that population level interventions could reduce 50-90$\%$ of strokes worldwide their implementation has been slow and far from universal \cite{owolabi2022primary}.  Targeting high-risk individuals is one of the strategies outlined in the European stroke guidelines \cite{piepoli_2016_2016}, where a predictive model is suggested to be used to determine the patients at high risk of a stroke. Such predictive models are becoming increasingly more popular with greater availability of large data sets.    
However, while predicting and reducing individual risk is important, it is also important to understand how society might be impacted by a the implementation of a high-risk individual level intervention. Knowing how a high-risk strategy might lower rates of strokes or reduce stroke severity across the population could be essential in public health planning.  

One method to assess the impacts of an intervention for high risk individuals on the population burden of stroke is to use an agent-based model. 
Agent-based models are a type of computer simulation made up of agents who can interact with each other and with their environment. Agents have a set of characteristics that can be used to guide their behaviour or health outcomes such as age, sex, or socioeconomic status \cite{gilbert2019agent}.  One advantage of agent-based models is their ability to combine a population approach with an individual approach. By modelling each agent's individual risk of stroke, we can then see how changes in their behaviours will impact the population burden of stroke. For example, we can simulate the impact of a set of agents reducing their stroke risk on the overall population burden of stroke or we can simulate the spread of healthy behaviours across a network of agents. 

In recent years the use of agent-based models for non-communicable diseases is increasing \cite{tracy2018agent}. However, the majority of agent-based models for non-communicable diseases focus on obesity, diabetes or health related behaviours such as physical activity, diet, smoking or alcohol consumption \cite{tracy2018agent,nianogo2015agent}. There are a few agent-based modelling studies that involve stroke  \cite{alassadi2021agent,hunter2022simulating,al2018agent,alassadi2022population}, however, these models all focus on the logistics of patients arriving at the hospital or delays in seeking treatment.  Models concerning agent behaviours that might increase stroke risk, are more broad and look at how changes in risk factors at the population level might impact the prevalence of cardiovascular disease \cite{li2014assessing,garney2022evaluating}. Thus there is a gap in the literature for a comprehensive stroke agent-based model that follows a patient on their stroke journey, developing stroke through a set of risk factors, seeking treatment for a stroke and determining the severity of the stroke based on not only the patients characteristics but their treatment delay as well.  

We propose to use an agent-based model that follows agents on their stroke journey from prevention through the acute phase of stroke and recovery to assess the impacts of a stroke interventions at the population level. The model can be used to assess the impacts of both population and high risk strategies, however, in this paper we present an example that shows the population level impacts of a high risk strategy.  Specifically,  the potential impact on the burden of stroke on society of a health conversation when a patient reaches certain ages that informs a patient of their stroke risk and leads to patients with high risk reducing their stroke risk.  In the next section we describe the model, we then discuss the experiments and finally the results.

\section{Methods}

\begin{figure}[t]
  \centering
  \includegraphics[width=0.4\textwidth]{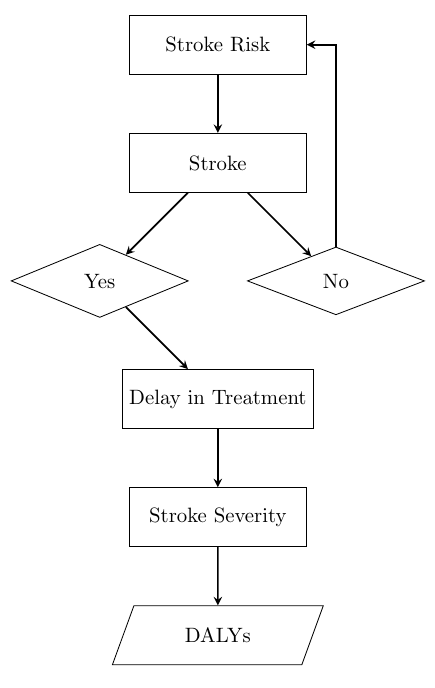}
  \caption{Figure showing an overview of an agent's journey within the model.}
  \label{fig:model}
\end{figure}

Here we provide a brief description of the model used and discuss the experiments that are run to assess the impacts of introducing a health conversation to make agents aware of their risks. The model used for our study is an agent-based model that is designed to simulate the disability adjusted life years (DALYS) due to stroke over a ten year period within a population.  The model is made up of a population of agents, each with their own set of characteristics.  These characteristics determine the agents’ risk of having a stroke and based on that stroke risk, each time step or day in the model some agents will have a stroke.  The model allows for agents to adjust their stroke risk by changing their modifiable risk factors (blood pressure, BMI, and smoking). If an agent has a stroke, the severity of that stroke will be determined and the severity will be used to determine the agent’s disability post-stroke.   Figure \ref{fig:model} provides an overview of what happens to each agent wihtin the model. Agent-based models are discrete models that run on time steps, in our model each time step equates to one day. The next sections describe the different components of the model in greater detail.  

\subsection{Agents}
Agent-based models at their core are driven by the agents. The agents in this model represent a person in a synthetic population that was created to match Irish demographics. The population is created using Irish Census Data and is scaled down so that each agent represents 100 people. This is done to save computing power and time.  Agents are matched to the Irish population so that in each county there are the correct proportions of agents of each gender, age, and employment status in the model. Agents household structures are also matched so that there are the right proportions of single households, couple households, and households with children in each county.  As we are focusing on stroke risk in the model we restrict the population in the model to only those over the age of 35. This is to match with the age ranges covered by the stroke risk prediction model discussed in Section \ref{section:risk}. There are a total of  22,119 agents in the model.  

After creating our agents, we then assign values for the stroke risk factors to each agent. An agent needs to be assigned a value for their systolic blood pressure, diastolic blood pressure, BMI, if they have diabetes, if they have atrial fibrillation, if they smoke and if they smoke how many cigarettes per day do they smoke. To assign a blood pressure to the agents, we separate agents into the following age bands: 50-59, 60-69 and 70 plus. For each age band we find the average and standard deviation of the risk factor in TILDA (the Irish Longitudinal Study on Aging) data \cite{tilda_irish_2016} and each agent samples a value for the risk factor from a normal distribution fitted to the TILDA parameters for their age category.  To determine which agents have diabetes, atrial fibrillation or smoke we determine the percent of people in an age group in TILDA with the risk factor and assign that percent of agents in the age group the risk factor.  TILDA data does not have information on the number of cigarettes smoked per day so to estimate this for the agents we use data from the Framingham Heart Study \cite{Framingham} and find the average number cigarettes smoked per day for smokers in each age group. 

\subsection{Stroke Risk} \label{section:risk}
To calculate stroke risk for each agent we use the age specific stroke risk models that were presented in \cite{hunter_age_2022} and combined as a weighted ensemble model in \cite{herrgaardh2022digital}.  The models are logistic regression models created to predict an individual's risk of primary ischemic stroke in five years. The age-specific models were created to account for the non-proportionality of stroke risk factors by age that can lead to over or under predictions in stroke risk \cite{hunter2023determining}.   

At the first day of each year in the model, the agents determine their 5-year primary stroke risk using the weighted ensemble model. As we run daily time steps in the model, the five-year stroke risk needs to be reduced to a daily stroke risk. To do this, we make the assumption that an individual's stroke risk is spread evenly throughout the five years. Once we make this assumption, we then divide the five year stroke risk by 1,826 (the number of days in five years) to get our daily absolute stroke risk.

\subsection{Stroke Occurrence}
Each day in the model, the absolute risk assigned to the agent using the stroke risk model described in Section \ref{section:risk} is used to determine if the agent had a stroke in the time step. For each agent who has not already had a stroke a random uniform number between 0 and 1 is drawn and if that random number is less than the agent's absolute risk of stroke the agent will have a stroke. 

If an agent has a stoke, their delay time in reaching the hospital will be determined. Determining the delay in reaching the hospital is important in determining stroke outcomes as rapid care has been associated with improved case-fatality rates \cite{mcgurgan_acute_2021,parry-jones2019}. Stroke is considered a neurological emergency and requires fast diagnosis and treatment to minimise the short and long-term health impacts \cite{mcgurgan_acute_2021}. To determine the delay in an agent reaching the hospital we use data from the Irish national audit of stroke from 2019 \cite{nationalstroke}, that provides a  distribution of the time from stroke symptom onset to hospital arrival in Ireland. Table \ref{table:delay} shows the percent of strokes and the delay time in reaching the hospital according the the Irish national audit of stroke from 2019.    

\begin{table}
\caption{Time from onset of stroke to hospital arrival from \cite{nationalstroke} }\label{table:delay}

\begin{tabular}{|l |c |}
\hline Hours & Percent of Strokes \\
\hline

0 - 3 &  49\\
3 - 4.5& 10 \\
4.5 - 12  &  20  \\
12 plus & 21 \\
\hline
\end{tabular}

\end{table}

To assign agents a delay they are first assigned a value between 0 and 1. If that number is less than or equal to 0.49, the agent will have a delay time between 0 and 3 hours (the exact delay for each agent is determined using a random normal distribution with mean 1.5 and standard deviation 0.75), if the number is greater than 0.49 but less than or equal to 0.59 the agent's delay time will be between 3 and 4.5 hours (determined using a random normal distribution with mean of 3.75 and standard deviation of 0.375), if the number is greater than 0.59 and less than o equal to 0.79 the delay time will be between 4.5 and 12 hours (determined using a random normal distribution with mean of 8.25 and standard deviation of 1.875), if the number is greater than 0.79 the agent will have a greater than 12 hour delay in getting to the hospital (determined by a random normal distribution with mean of 15 and standard deviation of 1).  

Once the delay time to the hospital has been determined for each agent with a stoke the outcome severity of the stroke is determined.  Agents can either have no disability, mild disability, severe or moderate disability or die from the stroke.  The percent of strokes in each level of severity is taken from the Irish national audit of strokes in 2019: 19$\%$ of ischemic strokes have no disability, 35$\%$ have mild disability post stroke, 37$\%$ have moderate to severe disability and 9$\%$ die from stroke \cite{nationalstroke}.  The percentages are adjusted for the agent's delay in arrival time to the hospital based off of results found in \cite{naganuma_early_2009}. The study shows that if the patient had an delay in arrival time of less than 3 hours the odds ratio for an mRS \footnote{The modified Rankin Scale is a clinical scale for assessing neurological disability. It ranges from 0 (no symptoms)  to 6 (dead).} less than or equal to 1 was 1.66 and for greater than equal to 2 was 1.73 with patients who had a delay of greater than or equal to 8 hours as the reference.  For patients who had a delay in arrival time between 3 and 8 hours the odds ratio for an mRS less than or equal to 1 was 1.15 and for greater than or equal to 2 was 0.98 with patients who had a delay of greater than or equal to 8 hours as the reference. 

Disability weights are also assigned to each stroke category based on the disability weights modified for western Europe in \cite{stouthard_disability_2000}.  Based on the protocol, if an agent has no disability they will have a disability weight of 0, if the agent has a mild disability they will have a disability weight of 0.35 and if they have a moderate/severe disability they will have a disability weight of 0.7. 

\subsection{Interventions}

The model allows agents to reduce their stroke risk in two scenarios: If they find out they have a high risk through a health conversation with their GP. If their family member has found out they have a high risk of stroke. 

\subsubsection{Health Conversations}
The health conversations in the model are aimed to mimic the health conversations that are done in Sweden when people reach a certain age where people are made aware of their health risk \cite{herrgaardh2022digital}.  Agents in the model will have  a conversation with their GP at the ages of 50, 60, 70, 80, and 90.  Agents will then know their stroke risk and if it is high they will undertake measures to reduce their risk.  If the agent smokes they will stop smoking, if their BMI is higher than the average BMI in the population, they will reduce their BMI by half of the population standard deviation of BMI.  They will also reduce their systolic and diastolic blood pressure by one tenth of a standard deviation of the population average blood pressure. 

\subsubsection{Family Members with High Risk}

In the model agents are also able to reduce their risk if a family member has been identified as having a high stroke risk after a health conversation.  If family members are reducing their risk they will reduce risk in the same way that agents reduce risk after a health conversation. This is included in the model to mimic family members supporting those with high stroke risk and participating in preventative measures.

\subsection{Estimating DALYs}

The model estimates the burden of stroke on society by calculating the disability adjusted life years (DALYs). DALYs are a measure that is used to quantify the burden of disease on the population and was developed for the Global Burden of Disease Study \cite{murray_understanding_1997}. They were developed as a measure that would quantify the magnitude of deaths and disability and a measure for cost-effectiveness analysis that would make the ethics of quantifying health more transparent.   

DALYs are calculated in the model using the formula  Years of Life Lost (YLL) plus Years Lost to Disability (YLD) \cite{devleesschauwer_calculating_2014}.  For agents who are deemed to have died from stroke their years of life lost will be calculated by taking the difference from their current age and their life expectancy. To calculate life expectancy we use the Irish Census life tables \cite{centralstatisticsoffice_2017}.  For agents who have not died from a stroke their YLD will be calculated by taking the difference between their age and their life expectancy and multiplying that difference by the disability weight that was determined when they had a stroke.  The population DALYs are then calculated by summing the YLD and YLL across all agents who have had a stroke and the average DALYs for the population is calculated by dividing the total DALYs by the number of agents who have had a stroke.

\subsection{Schedule}

At each model time step the following occurs: The model will assess the day of the year based on the number of time steps that have already occurred in the model run.  If the model moves to the next year, agents will increase their age by one and decrease their remaining life expectancy by 1.  Agents will then find their new stroke risk based on their updated age. If the agent is 50, 60, 70, 80 or 90 they will find out their stroke risk. If their stroke risk is high, absolute risk greater than 0.1, they will reduce their risk. If family members of those with high risk are reducing their risk, they will also reduce their risk.  At each time step of the model, agents will determine if they have a stroke. If they have a stroke they will seek care and determine the severity of their stroke.  DALYs will then be calculated.

\subsection{Output}
We look at two different outputs for the model. One is the number of total strokes in the population over a ten year period of time. This is done to determine if the high-risk intervention implemented (i.e, the health conversation) has had an impact on the number of strokes. The next output we look at is total DALYS across the ten years. This gives us an idea of the total burden of stroke in the population.

\subsection{Experiment}

We run an experiment using the model described in the previous section to understand how the health conversations might impact the burden of stroke on society. We look at two scenarios, the first where only agents who have been informed of their high risk at a health conversation reduce their risk and the second where agents who have been informed of a health conversation and their family members reduce their risk. Both scenarios are compared to a baseline model run where there is no reduction in risk. Because agent-based models are stochastic and each run will produce different results, the models are run 1,000 times to account for the variability in model output.  

We run each scenario for the equivalent of ten years in the model. For each scenario we look at two different measures of the impact of stroke on the population: the total number of strokes that have occurred, and the total DALYs and present the percent difference between the baseline scenario (no interventions or reduction in risk). We find these measures by taking an average across the 1,000 runs.  We look at the percent difference and not the magnitudes because the simulation we are running has a reduced population compared to the real population of Ireland. Thus the magnitude of  DALYs will be smaller in the simulation than what would be expected in the real world. The percent differences will give an estimate of the reduction in total DALYs in the population.  To test if the differences we see between the scenarios and the baseline are statistically significant we use student's t-test.

\section{Results}

In this section we discuss the results from the experiment run to look at the impact of agents learning about their stroke risk. We first look at the number of strokes that occur in each scenario, and the total DALYs from each scenario.  

\subsection{Strokes}

In an initial comparison of the reduction in stokes and costs of strokes due to patients understanding their stroke risk, the number of strokes over the ten year period is shown in Table \ref{table:stroke}. The table also shows the percent difference between the scenarios and the baseline and if the difference is statistically significant at the 95$\%$ level. We can see that there is a reduction of approximately 10 strokes over the 10 years compared to the baseline in scenario 1 and scenario 2 that equates to about a 1.4$\%$ reduction in strokes from the baseline to scenario 1 and a 2.1$\%$ reduction in strokes from the baseline to scenario 2.  We also see a statistically significant difference between scenario 1 and scenario 2.  

\begin{table}
\caption{Percent difference in strokes for the two scenarios compared to the baseline. }
\label{table:stroke}
\begin{tabular}{|l |l| l |}
\hline & Strokes & Percent Difference \\
\hline

Baseline & 551 & -\\
Scenario 1 &543 & -1.43$\%$ * \\
Scenario 2 & 539 & -2.12$\%$ *\\

\hline
\end{tabular}

\end{table}

\subsection{DALYs} 
Looking at the impact of the different scenarios on DALYs, Table \ref{table:DALYs} shows the total DALYs for ten years the model runs for the baseline, and the two scenarios. The table also shows the percent change between the scenarios and the baseline and indicates if the difference between the scenario and the baseline is statistically significant at a 95$\%$ level. Introducing health visits, where agents with high stroke risk reduce their risk results in a 1.16$\%$ reduction in DALYs. When the family members of agents with high risk also reduce their stroke risk DALYs decrease by 1.37$\%$.   We, however, do not see a statistically signficant difference in total DALYs between senario 1 and scenaro 2.

\begin{table}
\caption{Percent Difference in Total Population DALYs for the two scenarios compared to the baseline }
\label{table:DALYs}
\begin{tabular}{|l |l |l | }
\hline & DALYs & Percent Difference  \\
\hline
Baseline & 17344.6 & -\\
Scenario 1 & 17142.8  & -1.16$\%$ * \\
Scenario 2 & 17107.0 & -1.37$\%$ *\\
\hline
\end{tabular}

\end{table}

\section{Conclusion}

The results of our model show that by making agents aware of their stroke risk and giving agents the ability to reduce their risk if it is high, the overall number of stroke and the burden of stroke (assessed by total DALYs in the population) will be reduced. Additionally, if the risk reduction behaviours of the agents with high stroke risk spread to there family members we see an even further reduction of stoke. Although the reduction in strokes here is small, only 1.4 to 2.1$\%$, even a 2$\%$ decrease in the number of strokes would save lives. In 2019, there were approximately 5,550 stroke cases admitted to Irish hospitals, of which 86$\%$ were ischemic strokes \cite{nationalstroke}. Thus a 2.1$\%$ decrease in total strokes would result in an approximate reduction of 100 stroke cases in Ireland. That would equate to 100 people who would not have the long term impacts of a stroke. Similarly, the reduction in DALYs calculated by the model is small, just over a 1$\%$ decrease, but even in the model with a reduced population size this equates to over 200 Disability Adjusted Life Years saved. Suggesting that if individuals are aware of their stroke risks and act to reduce those risks there is a potential to save hundreds of healthy life years.   

The study, however, has a number of limitations. There are several assumptions made in model creation that lead to the study limitations. For example, in the model agents are assigned risk factors at model setup and with the exception of age that increase every year in the model, risk factors only change if agents reduce their risk.  In reality, many risk factors such as blood pressure or BMI will change over time.  This could lead to lower stroke cases in our model as we are not capturing the change in risk factors as agents age.  We also estimate DALYs based off of Irish data and give every agent an equally likely chance to have severe, moderate or mild outcomes based on the randomly generated delay time in arriving at the hospital. However, these wait times are likely not random with those living closer to hospitals more likely to have lower delay times and those living on there own likely to have longer delay times. Additionally, agents only reduce their risk if they are notified of their high risk at a health conversation or if they have a family member notified of high risk and they only reduce their risk once. In reality, agents might continuously reduce their risk, reduce their risk in one way, for example quitting smoking but not in others such as lowering BMI, or might not reduce their risk at all deciding to not change their lifestyle.  Future work will include looking at these other methods of reducing risk or looking at how risk reduction might spread through a network if someone has a stroke with severe outcomes. 

  While there are improvements that could increase the realism in the model, the work presented here is a step towards combining the fields of population health and personalized medicine in order to learn more about the health of the population and how individual interventions might impact society.  In this paper we have done an experiment to look at the potential impact of health visits where agents learn their stroke risk and how to reduce that risk.  However, the method we have used here is novel in that we take  an agent population and use machine learning stroke risk prediction models to estimate agents' risks and determine which agents will have a stroke.  Then based on this level of stroke incidence we determine the total DALYs  in the population. Other population interventions could be implemented to determine how they impact stroke incidence and the burden of stroke on society, for example, understanding how continuous risk stratification might impact outcomes, how a targeted prevention campaign to stop smoking, how creating more targeted stroke centers might impact the economic burden of stroke on society or how neighbourhood level impacts such as access to outdoor spaces or healthy food could impact the burden of stroke on society.  The simulation allows us to assess these interventions before they are put into place and understand how they might impact different areas of society.  

\subsubsection{Funding} 
 This work was partly supported by the PRECISE4Q  project funded by the European Union’s Horizon 2020 research and innovation programme under grant agreement No. 777107,  the STRATIF-AI project funded by EU's Horizon Europe research and innovation programme under grant agreement No. 101080875 and by the ADAPT Centre for Digital Content Technology funded under the SFI Research Centres Programme (Grant 13/RC/2106$\_$P2) and was co-funded under the European Regional Development Funds.

\subsubsection{Acknowledgements} 
This manuscript was prepared using FRAMCOHORT, GEN3, and FRAMOFFSPRING research materials obtained from the NHLBI Biologic Specimen and Data Repository Information Coordinating Center and does not necessarily reflect the opinions or views of the FRAMCOHORT, GEN3, FRAMOFFSPRING, or the NHLBI. \\
The Irish Longitudinal study on Ageing (TILDA)" and also ISSDA, in the following way: “Accessed via the Irish Social Science Data Archive - www.ucd.ie/issda
%
%
%
\bibliographystyle{splncs04}
\bibliography{StrokeDalysABM.bib}

\end{document}